# Kainate receptor modulation by Neto2

Lingli He[1,3]*, Jiahui Sun[4,5]*, Yiwei Gao[1,3]*, Bin Li[1,3], Yuhang Wang[1,3], Yanli Dong[1,3], Weidong An[1,3], Hang Li[1,3], Bei Yang[1,3], Yuhan Ge[4,5], Xuejun Cai Zhang[1,3]**, Yun Stone Shi[4,5,6,7]**, Yan Zhao[1,2,3]**

[1] National Laboratory of Biomacromolecules, CAS Center for Excellence in Biomacromolecules, Institute of Biophysics, Chinese Academy of Sciences, Beijing 100101, China

[2] State Key Laboratory of Brain and Cognitive Science, Institute of Biophysics, Chinese Academy of Sciences, 15 Datun Road, Beijing, 100101, China

[3] College of Life Sciences, University of Chinese Academy of Sciences, Beijing 100049, China

[4] Ministry of Education Key Laboratory of Model Animal for Disease Study, Model Animal Research Center, Medical School, Nanjing University, Nanjing 210032, China

[5] State Key Laboratory of Pharmaceutical Biotechnology, Department of Neurology, Affiliated Drum Tower Hospital of Nanjing University Medical School, Nanjing University, Nanjing 210032, China

[6] Institute for Brain Sciences, Nanjing University, Nanjing 210032, China

[7] Chemistry and Biomedicine Innovation Center, Nanjing University, Nanjing 210032, China

* These authors contribute equally to this project.

** Correspondence emails: zhaoy@ibp.ac.cn (Y.Z.), yunshi@nju.edu.cn (Y.S.S.), and zhangc@ibp.ac.cn (X.C.Z.)

## Abstract

Glutamate-gated kainate receptors (KARs) are ubiquitous in the central nervous system of vertebrates, mediate synaptic transmission on post-synapse, and modulate transmitter release on pre-synapse. In the brain, the trafficking, gating kinetics, and pharmacology of KARs are tightly regulated by Neuropilin and tolloid-like proteins (Netos). Here we report cryo-EM structures of homo-tetrameric GluK2 in complex with Neto2 at inhibited and desensitized states, illustrating variable stoichiometry of GluK2-Neto2 complexes, with one or two Neto2 subunits associate with the GluK2. We find that Neto2 accesses only two broad faces of KARs, intermolecularly crosslinking the lower-lobe of ATD$^{A/C}$, upper-lobe of LBD$^{B/D}$, and lower-lobe of LBD$^{A/C}$, illustrating how Neto2 regulates receptor-gating kinetics. The transmembrane helix of Neto2 is positioned proximal to the selectivity filter and competes with the amphiphilic H1-helix after M4 for interacting with an ICD formed by the M1-M2 linkers of the receptor, revealing how rectification is regulated by Neto2.

Kainate receptors are a class of ionotropic glutamate receptors, activated by neurotransmitter glutamate[1-3]. They are not only located in the post-synapse to mediate excitatory neurotransmission in many brain regions[4,5], but also appear in the pre-synapse to modulate transmitter release on both excitatory and inhibitory synapses[6,7]. Netos are single-pass transmembrane proteins with an extracellular domain containing two C1r/C1s-Uegf-BMP domains (CUB1 and CUB2), and a low-density lipoprotein class A domain (LDLa). These proteins have been identified as an auxiliary component of native KARs and significantly affects KARs trafficking, gating, and pharmacology[8-11]. More specifically, the Neto2 modulates KARs gating by slowing deactivation and desensitization, accelerating recovery from desensitization, and attenuating polyamine block of calcium-permeable KARs[12,13]. Despite recent progress in structural study of isolated KARs[14-18], molecular basis of regulatory roles of Netos remain unclear. Here, we show the architectures of GluK2-Neto2 complex in the antagonist-bound closed-state and agonist-bound desensitized-state, illustrating interactions and stoichiometry between GluK2 and Neto2, and the modulation mechanism of Neto2 on the GluK2 receptor gating and pore properties. Moreover, a more complete pore domain, including a detailed structure of the selectivity filter (SF), is provided in our structures.

## Cryo-EM analysis of GluK2-Neto2 complex

To investigate the structural basis for modulation of GluK2 gating by Neto2, we co-expressed both the full length GluK2 (with Gln at the Q/R site) and Neto2 in HEK293 cells and purified the complex (Extended Data Fig. 1)[19]. Our cryo-EM studies generated three distinct assemblies of the antagonist-bound GluK2-Neto2 complexes, including GluK2 bound with one Neto2 (GluK2-1×Neto2) or two Neto2 (GluK2-1×Neto2) (Fig.1a-1b and Extended Data Fig. 2a). The third type of complexes features a disrupted LBD dimer on one side and an intact LBD dimer on the other side, the latter of which is also bound with one Neto2 subunit, and thus this complex is denoted as GluK2-1×Neto2$^{asymLBD}$ (Fig. 1c and Extended Data Fig. 2a). The GluK2-1×Neto2, GluK2-2×Neto2, and GluK2-1×Neto2$^{asymLBD}$ complexes were determined at 4.2-Å, 6.4-Å, and 4.1-Å resolutions (Extended Data Figs. 2b–2j), respectively. The LBD-TMD focused classification and refinement of the GluK2-1×Neto2 complex yield a 3.9-Å map with more density features (Extended Data Figs. 2k-2m and 3a-3b). Detailed data processing and model building are described in the Methods section.

## Architectures of the GluK2-Neto2 complex

The overall structure of Neto2 is about 140 Å in height, composed of four domains including CUB1, CUB2, LDLa, and TM1 (Fig. 1d). There is a horizontal helix, termed as $\alpha1$, right before TM1 of Neto2 (TM1$^{Neto2}$) (Fig. 1d). The CUB1, CUB2, and LDLa domains are closely packed together. The linker between LDLa and $\alpha1$-helix was missing probably due to conformational heterogeneity (Fig. 1d). On the GluK2 side, the model of both GluK2-1×Neto2 and GluK2-2×Neto2 complexes reveals that the M3 gate is closed and degree of LBD clamshell closure is nearly identical to the isolated GluK2 LBD structure bound with LY466195 (PDB ID: 5KUH), confirming that the complex is stabilized at the antagonist-bound inhibited state (Fig. 1e and Extended Data Fig. 4a). In the case of the GluK2-1×Neto2 complex, the distance between the ATD and LBD layers at the Neto2-bound side is substantially shortened compared with the non-Neto2 bound side, thus breaking from the ideal overall two-fold symmetry of the receptor complex, evidenced by the ATD-LBD layers from the A and C positions are not superimposable with each other (Extended Data Fig. 4b). In addition, the overall structure of the GluK2-2×Neto2 complex also breaks two-fold symmetry (Extended Data Fig. 4b). In particular, the interaction between CUB1 and CUB2 on one side is more extensive than that on the other side.

The GluK2-1×Neto2$^{asymLBD}$ complex contains one disrupted LBD dimer, showing that LBD of the B subunit rotates approximate 72° relative to that two-fold related LBDs (Fig. 1f), which is also consistent with observations from GluK3 structures[16]. Interestingly, one extracellular domain of Neto2 connect ATD with LBD (Fig. 1c). We speculate that this domain is CUB1 as it has similar binding mode with GluK2 ATD (Extended Data Fig. 4c). Considering that the TM1$^{Neto2}$ helix remains present in the structure, the absence of CUB2 and LDLa in the EM map is probably due to conformational flexibility.

## Interactions between GluK2 and Neto2

On the extracellular side of both the GluK2-1×Neto2 and GluK2-2×Neto2 complexes, the N-terminal CUB1 of the Neto2 interacts with the lower lobe in the ATD layer of the A/C subunit of GluK2 (R2-lobe$^{A/C}$). In the LBD layer, the CUB2 and LDLa domains of Neto2 interact with the upper-lobe of subunit B/D (D1-lobe$^{B/D}$) and lower-lobe of subunit A/C (D2-lobe$^{A/C}$), respectively (Fig. 2a). This spatial arrangement of GluK2-Neto2 interactions leads to a stoichiometry between GluK2 and Neto2 of either 4:1 or 4:2, which is distinct from the 4:4 stoichiometry between the AMPA receptor and TARP$\gamma2$[20]. Taking a close look at the

GluK2-Neto2 interface, we found that some charged residues from GluK2 contact with the extracellular domains of Neto2. For instance, K216/R220 on the ATD layer and E723/R727 on the LBD layer potentially form electrostatic interactions with Neto2 (Figs. 2b and 2d; Extended Data Figs. 5a and 5c). Unlike a point-to-point interaction mode, CUB2 forms more extensive interactions with the Loop-1 (residues 448–453) (Fig. 2c). Although their sequence identity is low (Extended Data Fig. 5b), the Loops-1 from different isoforms adopt similar 3D conformations (Extended Data Fig. 5e). In the GluK2-1×Neto2$^{asymLBD}$ complex, one of the LBD dimers is disrupted (Fig.1f), consequently yielding a new interaction between CUB1 and D1-lobe$^{B/D}$. In particular, residues I780/Q784 of LBD$^{B/D}$ are involved in interaction with CUB1 (Fig. 2e). Based on sequence and structure alignment (Extended Data Fig. 5), these critical interaction sites are fairly conserved in amino-acid sequence and in 3D structure, except for ATD of GluK4 and GluK5 (ATD$^{GluK4/5}$). However, considering that Neto1 and Neto2 have different subunit-dependent regulation on GluK1 and GluK2[12,21], we speculate that these interactions determined in GluK2-Neto2 complex might not fully reflect in different combinations of GluK-Neto complex. Some changes of binding geometry or even new interactions might occur, which would profoundly alter regulatory effects of Neto on KARs. In the GluK4 and GluK5 subtypes, R220 is substituted to a negatively charged Asp at the equivalent site (Extended Data Fig. 5a), which would repel negatively charged residues from Neto2 and thus prevent CUB1 from interacting with ATD$^{GluK4/5}$. Interestingly, GluK5 is shown to specifically occupy the B/D positions of the channel[15,17,22]; thereby this unfavorable CUB1-ATD$^{GluK4/5}$ interaction would not prohibit the Neto2 from associating with and modulating the heteromeric GluK2/GluK5 complex, which is a major population in the brain[23]. Nevertheless, we were unable to nail down crucial residues of Neto2 involving in these GluK2-Neto2 interactions due to medium resolution of the Neto2 density.

To explore functional roles of these contacts, we substituted corresponding residues of GluK2 to Ala to disrupt the putative interactions with Neto2 (Extended Data Figs. 5a–5d). Strikingly, the I780A/Q784A double mutation profoundly accelerated desensitization, probably due to the mutation destabilizes the LBD dimer. Neto2 is able to slow desensitization of GluK2$^{I780A/Q784A}$. The other mutations showed little effect on gating kinetics of GluK2 alone (Fig. 2f and Extended Data Fig. 6a). However, these mutations substantially decrease effects of Neto2 on the GluK2 gating kinetics to different extents. In particular, Neto2 lost its function of regulating GluK2 upon introducing the K216A/R220A mutations (GluK2$^{ATD-2A}$) (Fig. 2f), in line with a previous study suggesting that negatively charged residues of CUB1 play vital roles in Neto2 modulation[24]. In addition, a triple point mutation K448A/D450A/K451A of GluK2 (GluK2[D1-3A]) was designed to disturb contacts between CUB2 and D1-lobe[B/D]. Compared with the WT GluK2-Neto2 complex, the GluK2[D1-3A]-Neto2 complex represents a faster desensitization kinetics (Fig. 2f). The Loop-1 of GluK2 was then substituted by the equivalent residues from GluK5 (GluK2[K5-Loop1]), and this mutant significantly reduce effects of Neto2 on desensitization kinetics. However, an E723A/R727A double mutation in the D2-lobe (GluK2[D2-2A]), which presumably disrupts the LDLa-D2-lobe interaction, showed a moderate decrease in slowing desensitization by Neto2 (Fig. 2f). The Neto2 is able to attenuate inward rectification of mutants discussed above (Extended Data Fig. 6b), suggesting these mutations could not damage association between GluK2 and Neto2.

## Ion conduction pore of KARs

The LBD-TMD focused refinement yielded a 3.9 Å-resolution map, which enabled us to build the most complete TMD model of kainate receptor so far (Figs. 3a and Extended Data Fig. 7). The pore profile reveals that the constriction sites are T652, A656, and T660 (Extended Data Fig. 7b), consistent with observations in previous studies[17,18]. Most importantly, the SF, consisting of the pore helix M2 (residues 608–620) and pore loop (residues 621–625), was determined for the first time (Extended Data Fig. 7b). Residues on M2 helices form extensive hydrophobic interactions with M1 and M3 helices from the same subunit and M3 helices from adjacent subunits (Extended Data Fig. 7c). Furthermore, our construct harbors a Q621 at the Q/R site located at the tip of the pore loop[25]. Its sidechain points upward to the central vestibule (Extended Data Figs. 7b and 7d), and is aligned with residues at Q/R site observed in AMPA receptor structures[26,27]. Previous investigations show that arginine at this site renders GluK2 $Ca^{2+}$-impermeable and attenuates polyamine block[28,29], presumably due to charge-charge repulsion. In addition, inside the SF, we determined a cation ligated by carbonyl oxygen groups from the four Q621 residues with bond lengths of 3.5-4 Å (Extended Data Figs. 7b and 7d). Given Gln at the Q/R site, we propose that this cation is either a calcium or sodium ion.

## Regulation of inward rectification

The long loops between M1 and M2 helices were resolved in the GluK2 map in the presence of Neto2. These loops extend into the cytosol and interact with each other, forming an intracellular cap domain (ICD) underneath the SF (Fig. 3b). Moreover, the amphiphilic H1-helix (residues 857-870) was built immediately after M4. The C-terminus of

H1 is positioned proximal to the C-terminus of M1 and M1-M2 loop from adjacent subunit. Moreover, TM1$^{Neto2}$ forms extensive hydrophobic interaction with TMD of GluK2. In particular, TM1$^{Neto2}$ and its N-terminal α1 helix directly contact with M1 of GluK2. One strip-like shape density was observed in between TM1$^{Neto2}$ and M4 of GluK2, and is supposed to be the hydrophobic tail of a lipid molecule and important for bridging indirect contacts between TM1$^{Neto2}$ and M4. On the intracellular side, the C-terminus of TM1$^{Neto2}$ competes with the H1-helix to interact with the C-terminus of M1 and the N-terminus of M2.

To investigate functional roles of the H1-helix and ICD, we deleted the H1-helix (GluK2$^{\Delta H1}$) and substituted the M1–M2 linker with GluA2 equivalent residues (GluK2$^{A2ICD}$). Interestingly, we found that these mutations do not change receptor gating kinetics in the absence of the Neto2. However, in the presence of Neto2, the GluK2$^{\Delta H1}$ and GluK2$^{A2ICD}$ mutants show slower and faster desensitization kinetics than the WT GluK2, respectively (Fig. 3c). We speculate that these diametrically opposite effects on receptor desensitization kinetics result from distinct ratio of the GluK2-Neto2 complex to the total receptors on the cell surface. Considering the H1-helix blocks the Neto2 binding site but the ICD directly interacts with Neto2 (Fig. 3b), more GluK2$^{\Delta H1}$ would be fully occupied by Neto2 after deleting the H1-helix, and in contrast, more Neto2 would dissociate from GluK2$^{A2ICD}$ once the M1-M2 loop is changed to the equivalent loop from GluA2.

In regarding to inward rectification, the GluK2$^{A2ICD}$ mutant is similar to the WT GluA2, but distinct from the WT GluK2, indicating the M1-M2 loop plays major roles in determining inward rectification property of glutamate receptors. The M1-M2 loop of GluA2 was replaced by that from GluK2 (GluA2$^{K2ICD}$) and resulted in a substantial decrease of the GluA2 rectification (Figs. 3d-3e and Extended Data Fig. 6c), further supporting that the M1-M2 loop is crucial for rectification. In the absence or presence of Neto2, the GluK2$^{\Delta H1}$ receptor shows comparable inward rectification as the WT GluK2. We hypothesize that Neto2 competes with H1-helix to stabilize receptor ICD, creating a physical barrier that prohibits polyamine from diffusing close to the SF, thereby eliminating both polyamine inhibition and thus inward rectification.

## Desensitized GluK2-Neto2 complex

We also obtained a 3.8-Å map of the desensitized GluK2-Neto2 complex (Fig. 4a, Extended Data Fig. 8 and Table 1), featuring GluK2 receptor binding with only one Neto2 subunit. The channel gate is closed, yet the LBD clamshell closure is nearly identical to the isolated LBD structure bound with kainate[30], indicating the complex is stabilized at a desensitized state (GluK2-1×Neto2$^{des}$) (Extended Data Fig. 9a). Interestingly, only CUB1 and TM1 of the Neto2 subunit are clearly visualized (Extended Data Fig. 9b). Consistent with desensitized KARs alone[18], the LBD layer of the GluK2-1×Neto2$^{des}$ complex shows disrupted LBD dimer and undergo remarkable rearrangement (Figs. 4a-4b and Extended Data Figs. 9c-9d). Structural comparison of LBD layers from GluK2-1×Neto2$^{des}$ complex and desensitized GluK2 alone shows that all of four D2 lobes connecting gating helices M3 are superimposable, adopt a pseudo four-fold symmetry, and are close to the channel central axis, suggesting that, in both the absence and presence of Neto2, GluK2 receptors share similar mechanisms to decouple agonist binding from channel opening (Fig. 4b). Strikingly, the D1 lobes at B-D positions clearly show displacement between these two complexes. In the presence of Neto2, the D1 lobe of B subunit approaches to ATD of the A subunit due to interactions with CUB1 of the Neto2 (Extended Data Fig. 9e). In this desensitized state, the CUB2 and LDLa domains were not well determined (Figs. 4a and Extended Data Fig. 9c), suggesting that the interactions observed in the inhibited state were disrupted upon desensitization. Taken all together, we speculate that, in the presence of Neto2, the inter-subunit connections of both ATD$^{A/C}$-LBD$^{B/D}$ and LBD$^{A/C}$-LBD$^{B/D}$, bridged by extracellular domains of Neto2, hinder rearrangement in the LBD layer upon desensitization, and thus slow down the desensitization kinetics. Interestingly, the ATD-CUB1 interaction is constitutively present in both inhibited and desensitized states. We hypothesize that the ATD-CUB1 interaction renders CUB1 an N-terminal anchor, while TM1 acts as a C-terminal anchor, constricting CUB2 and LDLa domains around the LBD layer during gating cycle and thus modulating channel gating, underlying the essential role of the ATD-CUB1 interaction in regulatory function of Neto2.

## Main references

1   Herb, A. *et al.* The KA-2 subunit of excitatory amino acid receptors shows widespread expression in brain and forms ion channels with distantly related subunits. *Neuron* **8**, 775-785 (1992).

2   Bettler, B. *et al.* Cloning of a novel glutamate receptor subunit, GluR5: expression in the nervous system during development. *Neuron* **5**, 583-595 (1990).

3   Traynelis, S. F. *et al.* Glutamate receptor ion channels: structure, regulation, and function. *Pharmacological reviews* **62**, 405-496 (2010).

4   Castillo, P. E., Malenka, R. C. & Nicoll, R. A. Kainate receptors mediate a slow postsynaptic current in hippocampal CA3 neurons. *Nature* **388**, 182-186 (1997).

5       Vignes, M. & Collingridge, G. L. The synaptic activation of kainate receptors. *Nature* **388**, 179-182 (1997).

6       MacDermott, A. B., Role, L. W. & Siegelbaum, S. A. Presynaptic ionotropic receptors and the control of transmitter release. *Annual review of neuroscience* **22**, 443-485 (1999).

7       Chittajallu, R. *et al.* Regulation of glutamate release by presynaptic kainate receptors in the hippocampus. *Nature* **379**, 78-81 (1996).

8       Zhang, W. *et al.* A transmembrane accessory subunit that modulates kainate-type glutamate receptors. *Neuron* **61**, 385-396 (2009).

9       Copits, B. A., Robbins, J. S., Frausto, S. & Swanson, G. T. Synaptic targeting and functional modulation of GluK1 kainate receptors by the auxiliary neuropilin and tolloid-like (NETO) proteins. *Journal of Neuroscience* **31**, 7334-7340 (2011).

10      Tang, M. *et al.* Neto1 is an auxiliary subunit of native synaptic kainate receptors. *Journal of Neuroscience* **31**, 10009-10018 (2011).

11      Tomita, S. & Castillo, P. E. Neto1 and Neto2: auxiliary subunits that determine key properties of native kainate receptors. *The Journal of physiology* **590**, 2217-2223 (2012).

12      Straub, C., Zhang, W. & Howe, J. R. Neto2 modulation of kainate receptors with different subunit compositions. *Journal of Neuroscience* **31**, 8078-8082 (2011).

13      Fisher, J. L. & Mott, D. D. The auxiliary subunits Neto1 and Neto2 reduce voltage-dependent inhibition of recombinant kainate receptors. *Journal of Neuroscience* **32**, 12928-12933 (2012).

14      Meyerson, J. R. *et al.* Structural mechanism of glutamate receptor activation and desensitization. *Nature* **514**, 328-334 (2014).

15      Mayer, M. L. Crystal structures of the GluR5 and GluR6 ligand binding cores: molecular mechanisms underlying kainate receptor selectivity. *Neuron* **45**, 539-552 (2005).

16      Kumari, J., Vinnakota, R. & Kumar, J. Structural and functional insights into GluK3-kainate receptor desensitization and recovery. *Scientific reports* **9**, 1-16 (2019).

17      Khanra, N., Brown, P. M., Perozzo, A. M., Bowie, D. & Meyerson, J. R. Architecture and structural dynamics of the heteromeric GluK2/K5 kainate receptor. *Elife* **10**, e66097 (2021).

18      Meyerson, J. R. *et al.* Structural basis of kainate subtype glutamate receptor desensitization. *Nature* **537**, 567-571, doi:10.1038/nature19352 (2016).

19  Honore, T. *et al.* Quinoxalinediones: potent competitive non-NMDA glutamate receptor antagonists. *Science* **241**, 701-703, doi:10.1126/science.2899909 (1988).

20  Zhao, Y., Chen, S., Yoshioka, C., Baconguis, I. & Gouaux, E. Architecture of fully occupied GluA2 AMPA receptor-TARP complex elucidated by cryo-EM. *Nature* **536**, 108-111, doi:10.1038/nature18961 (2016).

21  Copits, B. A. & Swanson, G. T. Dancing partners at the synapse: auxiliary subunits that shape kainate receptor function. *Nature Reviews Neuroscience* **13**, 675-686 (2012).

22  Kumar, J., Schuck, P. & Mayer, M. L. Structure and assembly mechanism for heteromeric kainate receptors. *Neuron* **71**, 319-331 (2011).

23  Pinheiro, P. S. *et al.* Selective block of postsynaptic kainate receptors reveals their function at hippocampal mossy fiber synapses. *Cerebral cortex* **23**, 323-331 (2013).

24  Li, Y.-J. *et al.* Neto proteins regulate gating of the kainate-type glutamate receptor GluK2 through two binding sites. *Journal of Biological Chemistry* **294**, 17889-17902 (2019).

25  Köhler, M., Burnashev, N., Sakmann, B. & Seeburg, P. H. Determinants of Ca2+ permeability in both TM1 and TM2 of high affinity kainate receptor channels: diversity by RNA editing. *Neuron* **10**, 491-500 (1993).

26  Chen, S. *et al.* Activation and desensitization mechanism of AMPA receptor-TARP complex by cryo-EM. *Cell* **170**, 1234-1246. e1214 (2017).

27  Zhang, D., Watson, J. F., Matthews, P. M., Cais, O. & Greger, I. H. Gating and modulation of a hetero-octameric AMPA glutamate receptor. Nature 594, 454-458, doi:10.1038/s41586-021-03613-0 (2021).

28  Bowie, D. & Mayer, M. L. Inward rectification of both AMPA and kainate subtype glutamate receptors generated by polyamine-mediated ion channel block. *Neuron* **15**, 453-462 (1995).

29  Kamboj, S. K., Swanson, G. T. & Cull-Candy, S. G. Intracellular spermine confers rectification on rat calcium-permeable AMPA and kainate receptors. *The Journal of physiology* **486**, 297-303 (1995).

30  Nayeem, N., Mayans, O. & Green, T. Correlating efficacy and desensitization with GluK2 ligand-binding domain movements. *Open biology* **3**, 130051 (2013).

## Figure legends

### Figure 1. Architectures of the GluK2-Neto2 complexes

Cryo-EM reconstructions of the GluK2-1×Neto2 (**a**), GluK2-2×Neto2 (**b**), and GluK2-1×Neto2$^{asymLBD}$ (**c**) complexes, bound with DNQX. Subunits A/C and B/D of the GluK2 receptor are colored in blue and red, respectively. The Neto2 were colored in orange. The height of each complex is indicated. Cross sections of the LBD layer of corresponding complexes are shown in the middle in a smaller scale. (**d**) Architecture of the GluK2-2×Neto2 complex. N-linked carbohydrates are shown as "sticks". (**e**) Top-down view of the ATD-CUB1 layer, LBD-CUB2-LDLa layer, and TMD-TM1$^{Neto2}$ layer. (**f**) Conformational rearrangement of the LBD layer between the GluK2-1×Neto2 complex and GluK2-1×Neto2$^{asymLBD}$ complex.

### Figure 2. Extracellular interactions between GluK2 and Neto2

(**a**) Overall structure of the GluK2-1×Neto2 complex. Subunits A/C and B/D of GluK2 are colored in blue and red, respectively. Neto2 are shown in transparent orange surface. Three regions involved in GluK2-Neto2 interaction are highlighted, including ATD (R2-lobe) with CUB1 (**b**), LBD (D1-lobe) with CUB2 (**c**), and LBD (D2-lobe) with LDLa (**d**). Side-chains of critical residues for GluK2-Neto2 interactions are shown as sticks. The EM map is overlaid on the models of GluK2 and Neto2 as transparent surfaces, colored in grey (receptor) and orange (Neto2), respectively. (**e**) CUB1 and LBD interactions determined in the GluK2-1×Neto2$^{asymLBD}$ complex. (**f**) Analysis of the time constant of desensitization of the WT GluK2 and GluK2 mutants, in either the presence or absence of Neto2. Time constant ratios of recordings with or without Neto2 are indicated. Each symbol represents a single cell recording, and *n* value represents independent cells for statistical analysis. Significances were determined using two-sided unpaired t-test. ****, $P < 0.0001$; ns (GluK2$^{ATD-2A}$), not significant, $P = 0.3959$; *** (GluK2$^{I780A/Q784A}$), $P = 0.0002$. Data are mean ± S.E.M.

### Figure 3. Ion conduction pore of GluK2 and modulation mechanism of inward rectification.

(**a**) EM density map of the LBD and TMD layers of the GluK2-1×Neto2 complex. Subunits A/C and B/D of GluK2 are colored in blue and red, respectively. The Neto2 is colored in orange. N-glycans are colored in yellow. (**b**) The interactions between GluK2 and Neto2 at the TMD and ICD layers. (**c**) Quantification of the desensitization time constant of the WT GluK2, GluK2$^{\Delta H1}$, and GluK2$^{A2ICD}$, in the absence or presence of Neto2. Each symbol represents a single recording. Significances were determined using two-sided unpaired t-test. ****, P < 0.0001. (**d**) Normalized current-voltage relationship of GluK2, GluA2 or their mutants. The sample size is the same as panel **e**. (**e**) The rectification index (RI) is calculated via dividing the current amplitude at +60 mV by that at −60 mV. Each symbol represents a single recording. *n* value represents independent cells for statistical analysis. Significances were determined using two-sided unpaired t-test. ****, P < 0.0001. Not significant, ns. GluK2 vs. GluK2$^{\Delta H1}$, P = 0.2155; GluK2+Neto2 vs. GluK2$^{\Delta H1}$+Neto2, P = 0.1229. Data are mean ± S.E.M.

Figure 4. Architecture of the desensitized GluK2-Neto2 complex

(**a**) The cryo-EM reconstruction of the GluK2-1×Neto2$^{des}$ complex. Subunits A/C and B/D of the GluK2 complex are colored in blue and red, respectively. The fragments of Neto2 were colored in orange. Cryo-EM map of the CUB2 domain is extracted from the GluK2-1×Neto2$^{des}$ map, shown at a lower threshold (yellow transparent) and overlaid on the same map of a higher threshold. (**b**) Organization of the D2 lobe of the GluK2-1×Neto2$^{des}$ (red and blue) and desensitized GluK2 alone (grey). COMs of the lobes are depicted as black dots. Distances and angles are indicated.

## Methods

### Electrophysiology

Electrophysiological experiments were carried out using HEK 293T (Gibco, USA, not authenticated; mycoplasma negative) cells, which were cultured in a 37°C incubator supplied with 5% $CO_2$. cDNAs encoding GluK2 (UniProt ID: P42260) was subcloned into the vector pCAGGS-IRES-EGFP, and Neto2 (UniProt ID: C6K2K4) was subcloned into the vector pCAGGS-IRES-mCherry. GluK2 mutations was made by overlapping PCR and confirmed by Sanger sequencing. The total transfection system was 1 μg. When co-expression was carried out, the ratio between GluK2 and Neto2 was 1:1. The HEK 293T cells were transfected with lipofectomine2000 reagents (Invitrogen), expressed for 24 to 36 h, and then dissociated with 0.05% trypsin. After resuspended, cells were placed on poly-D-lysine pretreated slides, and electrophysiological records were performed 4 h later.

Receptor desensitization were recorded on outside-out patches excised from the transfected HEK 293T cells. The external solution was (in mM): 140 NaCl, 2.5 KCl, 2 $CaCl_2$, 1 $MgCl_2$, 5 glucose, and 10 HEPES (pH 7.4). Patch pipettes (resistance 3 to 5 MΩ) were filled with a solution containing (in mM): 130 KF, 33 KOH, 2 $MgCl_2$, 1 $CaCl_2$, 11 EGTA, and

10 HEPES (pH 7.4). Glutamate (10 mM) diluted into the external solution was applied for 500 ms with a theta glass pipette mounted on a piezoelectric bimorph [25] every 12 s. Glutamate induced currents were recorded with holding potential of -70 mV and fitted with a double exponential function $A = A_0*(f_1*exp(-t/\tau_f) + f_2*exp(-t/\tau_s)) + C$. In these functions $t$ is the time. The currents amplitude ($A$) starts at $A_0$ and decays down to steady state $C$. $f_1$ and $f_2$ are the fractions of respective components as percent ($f_1 + f_2 = 1$), and $\tau_f$ and $\tau_s$ are decay kinetics of fast and slow components. The weighted $\tau_{des}$ was calculated using the formula: weighted $\tau_{des} = f_1*\tau_f + f_2*\tau_s$.

The rectification of the receptors was analyzed using whole cell recordings on transfected HEK293T cells. The extracellular solution was the same as that used in patch recording. The glass pipettes (3 to 5 MΩ) filled with intracellular solution (in mM): 140 CsCl, 4 $MgCl_2$, 1 EGTA, and 10 HEPES, 4 $Na_2ATP$, 0.1 spermine (pH 7.2). Desensitization curves (10 mM glutamate for 200 ms) were recorded while the holding potential was elevated from -100 with a step of 20 mV to +100 mV. The RI (rectification index) was calculated using the relative current amplitudes at +60 mV and -60 mV: $RI = (I_{+60mV} - I_{0mV})/(I_{0mV} - I_{-60mV})$. All the currents were collected with an Axoclamp 700B amplifier and Digidata 1440A (MolecularDevices, Sunnyvale, CA, USA), filtered at 2 kHz, and digitized at 10 kHz for whole cell recordings and 100 kHz for outside-out patches. The current data were analyzed using Clampfit software.

### GluK2-Neto2 complex expression and purification

The full-length rat GluK2 cDNA sequence (UniProt ID: P42260) and rat Neto2 cDNA sequence (UniProt ID: C6K2K4) was cloned into pEG BacMam vector [31]. Specifically, GluK2 and Neto2 were fused with a PreScission protease cleavage site (SNSLEVLFQ/GP), and a C-terminal mCherry-Twin strep tag and a GFP-His$_8$ tag, respectively. An incidental point mutation, F107L (GluK2$^{F107L}$), was introduced at the ATD layer during the cloning. Electrophysiology experiment indicates that GluK2$^{F107L}$ does not profoundly alter channel desensitization kinetics. Moreover, the Neto2 exerts similar modulation on the WT GluK2 and GluK2$^{F107K}$ in slowing down desensitization (Extended Data Figs. 1a–1c). This variant was only used in the cryo-EM study, whereas the subsequent functional study was performed using the wild-type (WT) GluK2. Subsequently, bicistronic bacmid and baculovirus harboring the GluK2 and Neto2 genes were generated and P2 viruses were used to infect suspension HEK 293F (Gibco, USA; not authenticated; mycoplasma negative) cells at cell density of ~2×10$^6$. Cell culture was infected with 10% (v/v) Gluk2-Neto2 baculoviruses to initiate the transduction. Twelve hours after infection, 10 mM sodium butyrate and 20 μM DNQX were supplemented. Cells were harvested ~60 h post infection and stored at −80°C. We introduced antagonist DNQX (6,7-dinitroquinoxaline-2,3-dione, 20 μM) during the complex expression to reduce potential cell toxicity.

Cell pellets were resuspended in ice-cold buffer containing 20 mM HEPES (pH 7.5), 150 mM NaCl, 5 mM β-mercaptoethanol, and protease inhibitors Cocktail (Roche, Swiss). The membrane was collected by ultracentrifugation at 4°C (100,000 $g$ for 1 h). The GluK2-Neto2 complex was extracted with buffer containing 20 mM HEPES (pH 7.5), 150 mM NaCl, 5 mM β-mercaptoethanol, 20 μM DNQX, and 1% (w/v) digitonin for 3 h at 4°C. Insoluble material was removed by centrifugation (100,000 $g$, 1 h). The supernatant was filtered and passed through Streptactin Beads at 4°C. The beads were washed with 20 mM HEPES (pH 7.5),150 mM NaCl, 5 mM β-mercaptoethanol, 20 μM DNQX, and 0.08% (w/v) digitonin. Then, the GluK2-Neto2 complex was eluted with buffer containing 5 mM desthiobiotin. The fluorescent tags were cleaved by incubating with PreScission protease overnight at 4°C. The GluK2-Neto2 complex was further purified by gel filtration chromatography (Superose-6, 10/300) with a buffer containing 20 mM HEPES (pH 7.5), 150 mM NaCl, 5 mM β-mercaptoethanol, 20 μM DNQX, and 0.08% (w/v) digitonin. Peak fractions were pooled and concentrated to ~4.7 mg/ml for cryo-EM grids preparation. FSEC (fluorescence-detection size-exclusion chromatography) experiment confirmed formation of the GluK2-Neto2 complex (Extended Data Figs. 1d–1f).

Purification of kainite-bound desensitized GluK2-Neto2 complex was carried out following the same procedure of DNQX-bound inhibited complex without inhibitor supplemented during protein expression and purification. Agonist kainate was added to concentrated protein to achieve final concentrations of 5 mM immediately before cryo-EM grids preparation.

Cryo-EM sample preparation and data acquisition

Quantifoil R1.2/1.3 Cu 300 mesh grids were glow discharged for 60 s in $H_2$-$O_2$ condition. 2.5 µL of the GluK2-Neto2 complex at 4.7 mg/ml was applied to the grid followed by blotting for 2.0 s at 100% humidity and 4°C, and flash-frozen in liquid ethane using a Vitrobot Mark IV (Thermo Fisher Scientific, USA).

Grids were imaged with a 300 kV Titan Krios (Thermo Fisher Scientific, USA) or a 200 kV Talos Arctica (Thermo Fisher Scientific, USA), both equipped with a K2 Summit direct electron detector (Gatan, USA) and a GIF-Quantum energy filter. The slit width was set to 20 eV. A calibrated magnification of 105,000× was used, yielded a pixel size of 1.36 Å or 1.32 Å on images (Extended Data Table 1). The defocus range was set to between −1.2 and −2.2 µm. All movie stacks were collected by SerialEM under a dose rate of 9.1 ~ 9.4 e−/pixel/s with a total exposure time of 11.4 s, and dose-fractioned to 32 frames, resulting a total dose of 60 or 50 e−/Å$^2$.

## Single particle cryo-EM data processing

For DNQX-bound inhibited GluK2-Neto2 complex, a total of 5,448 movie stacks were collected, followed by motion correction using MotionCor2 [32] with 5 × 5 patches. Parameters of the contrast transfer function (CTF) were estimated using Gctf [33], followed by particle picking using Gautomatch and Topaz [34]. Particles were classified into eight classes using a GluK2$_{EM}$ density map (EMDB ID: EMD-8289 [18]) as the reference map in RELION [35]. Three resulting maps (classes 6, 7, and 8) displayed classic kainate receptor-like structural features, including ATD, LBD, and TMD. Moreover, one or two strings of globular densities were observed nearby GluK2. They appeared not to be an intrinsic part of GluK2 and were thus considered as soluble domains of Neto2. LBD densities in class 6 displayed an asymmetric organization, while their counterparts in classes 7 and 8 displayed a conventional dimer-of-dimers configuration. According to the different conformational states of LBD and stoichiometry of GluK2 and Neto2, the three classes were denoted as GluK2-1×Neto2$^{asymLBD}$ (class 6, 9.80%), GluK2-2×Neto2 (class 7, 5.09%), and GluK2-1×Neto2 (class 8, 6.42%), respectively. Particles were then imported to cryoSPARC for processing [36]. Non-uniform refinement gave rise to initial maps at 4.3 Å, 7.1 Å, and 5.1 Å, respectively. Two rounds of *Ab-initio* Reconstruction were subsequently conducted to remove junk particles, followed by another round of Heterologous Refinement using low-passed maps as references, at 7, 15, and 30 Å, respectively. The final round of non-uniform refinement generated 4.1-Å, 6.4-Å, and 4.2-Å full-length reconstructions according to the golden-standard *Fourier* shell correlation (GSFSC) criterion [37], illustrating hallmark features of glycosylation sites, LBD clamshells, and transmembrane helices. The LBD-TMD map of the GluK2-2×Neto2 complex is significantly improved by LBD-TMD focused refinement in the presence of C2 symmetry, generating a 5.6-Å map with more continuous transmembrane helices.

To improve the resolution of LBD-TMD of GluK2-1×Neto2 complex, focused classification and refinement were conducted. Masks were created using a GluK2-1×Neto2 map whose ATDs were manually erased. Masked 3D classification in RELION generated 3 classes [35]. Among them, only class 1 (28.2%) displayed continuous and clear TM densities. The final map of LBD-TMD was generated by non-uniform refinement in cryoSPARC with a LBD-TMD mask, which was reported at 3.9 Å according to the GSFSC criterion [37]. GluK2 LBD and TMD were well resolved in this new map with abundant features, including sidechain density for most aromatic residues, visible loops between LBD and TMD, and intact structures of both the SF and pore loop, which allowed us to build the TMD of GluK2. The linker between M2 and M3, for the first time, became visible and aided us to trace the backbone of this linker. This map also reveals two N-glycans inside linkers between the LBD and TMD layer, and these glycosylation sites facilitate sequence registration for S1-preM1 and S2-M4 linkers during model building.

For kainate-bound desensitized GluK2-1×Neto2 complex, a total of 2,957 movie stacks were collected, followed by motion-correction using MotionCor2, CTF estimation using Gctf, and particle picking using Gautomatch and Template Picker (cryoSPARC). The initial 3D classification without imposition of mask or symmetry generated 8 classes, 4 of which displayed kainate receptor-like structural features but with poor Neto2 density and were subjected to the second round of 3D classification in RELION, yielding 3 classes. The class 1 showed a density connecting ATD-LBD and clearly represent the GluK2-Neto2 complex. One round of Ab-initio Reconstruction and two rounds of Heterologous Refinement were then carried out in cryoSPARC to further clean up particles. The map quality was substantially improved. The final map was generated using Non-uniform Refinement, and the structure was determined at 3.8 Å according to GSFSC criterion [37].

Model building

Model building of the LBD-TMD part of GluK2 started with the 3.9-Å LBD-TMD map, containing receptor LBD-TMD and Neto2 CUB1-LDLa-TM1. The map clearly showed densities for most of aromatic residues and N-glycans, which allowed us to reliable build the model. The D1 lobe (residues 431–535 and 762–806), D2 lobe (residues 536–544 and 669–761) and TMD (residues 545–668 and 806–850) were extracted from the structure of *R. norvegicus* GluK2 (PDB ID: 5KUF [18]). The homology models of CUB2 and LDLa were prepared using the SWISS-MODEL utility based on homology structures with PDB IDs of 2WNO and 6H03, respectively [38,39]. All of these domains were fitted into the cryo-EM map as rigid bodies using UCSF Chimera software. The resulting model was then manually inspected and adjusted in COOT [40], followed by refinement against the cryo-EM map in real space using the phenix.real_space_refine utility [41].

The model of CUB1 of Neto2 was generated using SWISS-MODEL based on a homology model (PDB ID: 2QQL) [42]. The ATD layer of the GluK2 was extracted from a crystal structure of GluK2 (PDB ID: 5KUF [18]). Subsequently, the homology models of ATD and LBD-TMD of GluK2, CUB1, and CUB2-LDLa-TM1 of Neto2 were docked into cryo-EM maps of GluK2-1×Neto2, GluK2-2×Neto2, and GluK2-1×Neto2$^{asymLBD}$ complex as rigid bodies, generating corresponding complex structures. The resulting models were subsequently manually inspected and adjusted in COOT. Given the medium resolution of the maps of overall structure, the models of rigid body were refined against the cryo-EM map in real space using phenix.real_space_refine.

For GluK2-1×Neto2$^{des}$ complex, the ATD layer, D1-lobe, D2-lobe, TMD and CUB1 were extracted from the structure of GluK2-1×Neto2, which were docked into the cryo-EM map as rigid bodies using UCSF Chimera software. The resulting model was then manually inspected and adjusted in COOT, followed by refinement against the cryo-EM map in real space using phenix.real_space_refine. The model stereochemistry was evaluated using the Comprehensive validation (cryo-EM) utility in the PHENIX software package.

All figures were prepared with ChimeraX or PyMOL (Schrödinger, LLC) [43,44].

## Methods references

31   Goehring, A. *et al.* Screening and large-scale expression of membrane proteins in mammalian cells for structural studies. *Nature protocols* **9**, 2574-2585, doi:10.1038/nprot.2014.173 (2014).

32   Zheng, S. Q. *et al.* MotionCor2: anisotropic correction of beam-induced motion for improved cryo-electron microscopy. *Nature methods* **14**, 331-332, doi:10.1038/nmeth.4193 (2017).

33   Zhang, K. Gctf: Real-time CTF determination and correction. *Journal of structural biology* **193**, 1-12 (2016).

34   Bepler, T., Kelley, K., Noble, A. J. & Berger, B. Topaz-Denoise: general deep denoising models for cryoEM and cryoET. *Nature communications* **11**, 5208, doi:10.1038/s41467-020-18952-1 (2020).

35   Zivanov, J. *et al.* New tools for automated high-resolution cryo-EM structure determination in RELION-3. *elife* **7**, e42166 (2018).

36   Punjani, A., Rubinstein, J. L., Fleet, D. J. & Brubaker, M. A. cryoSPARC: algorithms for rapid unsupervised cryo-EM structure determination. *Nature methods* **14**, 290-

296 (2017).

37  Scheres, S. H. & Chen, S. Prevention of overfitting in cryo-EM structure determination. *Nature methods* **9**, 853-854, doi:10.1038/nmeth.2115 (2012).

38  Bordoli, L. *et al.* Protein structure homology modeling using SWISS-MODEL workspace. *Nature protocols* **4**, 1 (2009).

39  Briggs, D. C. *et al.* Metal ion-dependent heavy chain transfer activity of TSG-6 mediates assembly of the cumulus-oocyte matrix. *Journal of Biological Chemistry* **290**, 28708-28723 (2015).

40  Emsley, P. & Cowtan, K. Coot: model-building tools for molecular graphics. *Acta crystallographica section D: biological crystallography* **60**, 2126-2132 (2004).

41  Afonine, P. V. *et al.* Real-space refinement in PHENIX for cryo-EM and crystallography. *Acta Crystallographica Section D: Structural Biology* **74**, 531-544 (2018).

42  Appleton, B. A. *et al.* Structural studies of neuropilin/antibody complexes provide insights into semaphorin and VEGF binding. *The EMBO Journal* **26**, 4902-4912, doi:https://doi.org/10.1038/sj.emboj.7601906 (2007).

43  Goddard, T. D. *et al.* UCSF ChimeraX: Meeting modern challenges in visualization and analysis. *Protein Science* **27**, 14-25 (2018).

44  DeLano, W. L. Pymol: An open-source molecular graphics tool. *CCP4 Newsletter on protein crystallography* **40**, 82-92 (2002).

## Acknowledgments

We thank X. Huang, B. Zhu, X. Li, L. Chen, and other staff members at the Center for Biological Imaging (CBI), Core Facilities for Protein Science at the Institute of Biophysics, Chinese Academy of Science (IBP, CAS) for the support in cryo-EM data collection. We thank Prof. Nengyin Sheng for providing the cDNAs of GluK2 and Neto2. We thank Yan Wu for his research assistant service. This work is funded by Chinese Academy of Sciences Strategic Priority Research Program (Grant XDB37030304 to Y.Z. and Grant XDB37030301 to X.C.Z), National Key R & D Program of China (2019YFA0801603 to Y.S.S.), the National Natural Science Foundation of China (91849112 to Y.S.S., 31971134 to X.C.Z.), the Natural Science Foundation of Jiangsu Province (BE2019707 to Y.S.S.) and Fundamental Research Funds for the Central Universities (0903-14380029 to Y.S.S.).

## Author contribution

## Conflict of interest

All authors declare that there is no conflict of interest that could be perceived as prejudicing the impartiality of the research reported.

## Data availability

The three-dimensional cryo-EM density maps of the antagonist DNQX-bound GluK2-1×Neto2, LBD-TMD of GluK2-1×Neto2, GluK2-2×Neto2, GluK2-1×Neto2$^{asymLBD}$ complex and agonist kainate-bound desensitized GluK2-1×Neto2$^{des}$ complex have been deposited in the EM Database under the accession codes EMD-31462, EMD-31464, EMD-31463, EMD-31459 and EMD-31460, respectively. the cryo-EM map LBD-TMD of GluK2-2×Neto2 complexes have been deposited as an additional map under entry EMD-31463. The coordinates for the structures have been deposited in Protein Data Bank under accession codes 7F59, 7F5B, 7F5A, 7F56 and 7F57, respectively.

## Extended data figure legends

### Extended Data Figure 1. Functional study and purification of GluK2-Neto2 complex.

(**a**) – (**b**) Outside-out recording of the WT GluK2 and GluK2$^{F107L}$, in the absence or presence of Neto2. (**c**) Statistical analysis of the desensitization time constant of the WT GluK2 and GluK2$^{F107L}$, with or without Neto2 (GluK2, $n$ = 12, GluK2 + Neto2, $n$ = 14; GluK2$^{F107L}$, $n$ = 10; GluK2$^{F107L}$ + Neto2, $n$ = 10). Each symbol represents a single cell recording, and $n$ value represents biologically independent cells for statistical analysis. Significances were determined using two-sided unpaired t-test. ****, P < 0.0001. Similar results were reproduced from two independent experiments. Error bars stand for S.E.M. (**d**) Fluorescence-detection size-exclusion chromatography (FSEC) analysis of the co-expressed GluK2-mCherry (red) and Neto2-GFP (green). The experiments were repeated independently with more than three times with similar results. (**e**) Size-exclusion chromatography (SEC) profile of the purified GluK2-Neto2 complex. Fractions within the dashed lines were pooled for cryo-EM sample preparation. The experiments were repeated independently with more than three times with similar results. (**f**) Coomassie blue-stained SDS-PAGE gel of the pooled fractions. The gel was repeated three times from different batches of purification with similar results. The uncropped gel can be found in Supplementary Figure 1.

Extended Data Figure 2. Cryo-EM data analysis of GluK2-1×Neto2, GluK2-2×Neto2, and GluK2-1×Neto2$^{asymLBD}$ complex.

(**a**) Flowchart of cryo-EM data processing. A total of 5,448 movie stacks were collected and motion-corrected, followed by CTF estimation and particle picking. A representative motion-corrected micrograph of this dataset is shown here (Scale bar = 40 nm). The experiments were repeated three times with similar results. Particles were cleaned and classified using several rounds of 2D and 3D classifications, which generated 3 classes, representing GluK2-1×Neto2, GluK2-2×Neto2, and GluK2-1×Neto2$^{asymLBD}$, respectively. Particles were then submitted to further 3D classifications separately to improve resolutions. Focused classification and refinement of LBD-TMD were conducted on the particles of GluK2-1×Neto2 complex. Masks used in focused processing were overlaid on GluK2 map (green) as transparent grey surfaces alongside the arrows. (**b**, **e**, **h**, **k**) Angular distribution of the particles contributing the final reconstruction for GluK2-1×Neto2$^{asymLBD}$ complex (**b**), GluK2-2×Neto2 complex (**e**), GluK2-1×Neto2 complex (**h**), and LBD-TMD (**k**). The length of each spike indicates of the number of particles in the designated orientation. (**c**, **f**, **i**, **l**) Sharpened map of GluK2-1×Neto2$^{asymLBD}$ complex (**c**), GluK2-2×Neto2 complex (**f**), GluK2-1×Neto2 complex (**i**), and LBD-TMD (**l**), colored according to local resolution estimation. (**d**, **g**, **j**, **m**) The half-map (red) and model-map (black) *Fourier* shell correlation (FSC) of GluK2-1×Neto2$^{asymLBD}$ complex (**d**), GluK2-2×Neto2 complex (**g**), GluK2-1×Neto2 complex (**j**), and LBD-TMD (**m**).

Extended Data Figure 3. EM maps for transmembrane helices and the LBD-TMD.

(**a**) Transmembrane helices M1−M4, and the M1-M2 loop. EM maps are shown as transparent grey surfaces. Some sidechains are shown as sticks. (**b**) EM maps for LBD and TMD layers. CUB2 and LDLa of Neto2 are colored in orange. Receptor is colored in purple.

N-glycans and a lipid tail are shown in sticks.

Extended Data Figure 4. Structural comparison of ATD and LBD layers

(**a**) Superimposition of antagonist bound LBDs of 5KUH (grey) and GluK2-1×Neto2 (red). (**b**) Comparison of the $ATD^A$-$LBD^B$ and $ATD^C$-$LBD^D$ segments between subunit A (grey) and C (blue) of the GluK2-1×Neto2, GluK2-2×Neto2 and GluK2-1×Neto2$^{asymLBD}$ complexes, using LBD as a reference. The centers of masses (COMs) of the ATD R1/R2-lobe of subunits A and C are depicted as rectangles or triangles, respectively. The COMs of the LBD layer is marked as a circle. (**c**) Superimposition of ATD-CUB1 interactions between GluK2-1×Neto2$^{asymLBD}$ (grey) and GluK2-1×Neto2 (red, orange and blue).

Extended Data Figure 5. Sequence alignments and structural comparison of the KARs.

(**a**) – (**d**) Sequence alignments of the GluK members in *Rat norvegicus*, numbered according to full-length subunits. Secondary structures of GluK2 are marked above the sequence alignment. Dashes represent gaps. Conserved residues are shaded in grey. Residues which are involved in Neto2 interaction are indicated by triangle symbol. (**e**) Structural comparison of the LBD of GluK2 with GluK1 (3FUZ, green), GluK3 (3U92, cyan), GluK4 (5IKB, magenta), and GluK5 (7KS0, yellow), respectively. D1- and D2- lobes and Loop 1 are indicated.

Extended Data Figure 6. Representative desensitization and rectification traces

(**a**) Representative desensitization traces of GluK2 and mutants responded to 60 ms application of 10 mM glutamate were normalized and aligned to the peak. Superimposed responses of the receptor alone and the receptor-Neto2 complex were shown in black and green traces, respectively. (**b**) Normalized current-voltage relationship of GluK2 mutants in the absence and presence of Neto2. *n* value represents independent cells for analysis. (**c**) Representative desensitization traces and related statistical analysis (GluA2, *n* = 16; GluA2$^{K2ICD}$, *n* = 12). Each symbol represents a single cell recording, and *n* value represents biologically independent cells for statistical analysis. Significances were determined using two-sided unpaired t-test. Not significant (ns), P = 0.0755. No adjustments were made for multiple comparisons. Error bars stand for S.E.M. The H1-helix is composed of residues $^{857}$FCSAMVEELRMSLK$^{870}$ and removed in GluK2$^{\Delta H1}$ construct. The amino acid sequence of the ICD of GluK2 and GluA2 between M1-M2 helices are $^{587}$YEWYNPHPCNPDSDVVEN$^{604}$ and $^{570}$YEWHTEEFEDGRETQSSESTNE$^{591}$, respectively, which are involved in the ICD swapping constructs of GluA2$^{K2ICD}$ and GluK2$^{A2ICD}$. (**d**) I-V relationship for GluA2, GluK2 and related mutants. Desensitization curves (10 mM glutamate for 200 ms) were recorded at holding potential ranging from −100 to +100 mV in 20 mV increasement. Traces were normalized to the peak value at −100 mV.

Extended Data Figure 7. Interactions stabilizing the pore helix M2.

(**a**) EM density map of the LBD and TMD layers of the GluK2-1×Neto2 complex. Subunits A/C and B/D of GluK2 are colored in blue and red, respectively. The Neto2 protein is colored in orange. N-glycans are colored in yellow. (**b**) The ion conduction pore and its profile of the GluK2-Neto2 complex. Pore loops are colored in green. A cation ion is shown as a grey sphere, overlaid with corresponding EM density colored in marine. Q621, T652, A656, and T660 are shown in sticks. Constriction sites are indicated in the pore profile. (**c**) The TM helices of the GluK2 (red and blue) and the Neto2 (yellow) are shown as cartoon. The EM density of M2 helix are shown as transparent grey surface. Critical residues involved in interactions are shown as sticks. (**d**) "Top-down" view of M2 helices and the pore loops. Q621 residues are shown in sticks.

Extended Data Figure 8. Cryo-EM data analysis of GluK2-1×Neto2$^{des}$ complex.

(**a**) Flowchart of cryo-EM data processing. A total of 2,957 movie stacks were collected and motion-corrected, followed by CTF estimation and particle picking. A representative motion-corrected micrograph of this dataset is shown here (Scale bar = 40 nm). The experiments were repeated three times with similar results. Three-dimensional classification generated 8 classes, 4 of which displayed classical structures of kainate receptors. Another round of 3D classification was then performed, followed by *Ab-initio* Reconstruction and Heterologous Refinement to furtherly improve the quality of map. (**b**) Angular distribution of the particles contributing the final reconstruction of GluK2-1×Neto2$^{des}$ complex. The length of each spike indicates of the number of particles in the designated orientation. (**c**) Sharpened map of GluK2-1×Neto2$^{des}$ complex, colored according to local resolution estimation. (**d**) The half-map (red) and model-map (black) Fourier shell correlation.

Extended Data Figure 9. Conformational change of GluK2-Neto2 complex upon desensitization.

(**a**) Superimposition of agonist bound LBDs of 4BDM (grey) and GluK2-1×Neto2$^{des}$ (red). (**b**) Superimposition of ATD-CUB1-LBD interactions between GluK2-1×Neto2$^{asymLBD}$ (grey) and GluK2-1×Neto2des (red, orange and blue). (**c**) Organization of the D1 lobe of the GluK2-

1×Neto2$^{des}$ (red and blue) and desensitized GluK2 alone (grey). COMs of the lobes are depicted as black dots. Distances and angles are indicated. (**d**) The LBD rearrangement between GluK2-1×Neto2 and GluK2-1×Neto2$^{des}$ upon desensitization. (**e**) Displacement of LBD at B-position of GluK2-1×Neto2$^{des}$ complex (red, blue and orange) compared with desensitized GluK2 (5KUF, grey).

Figure 1

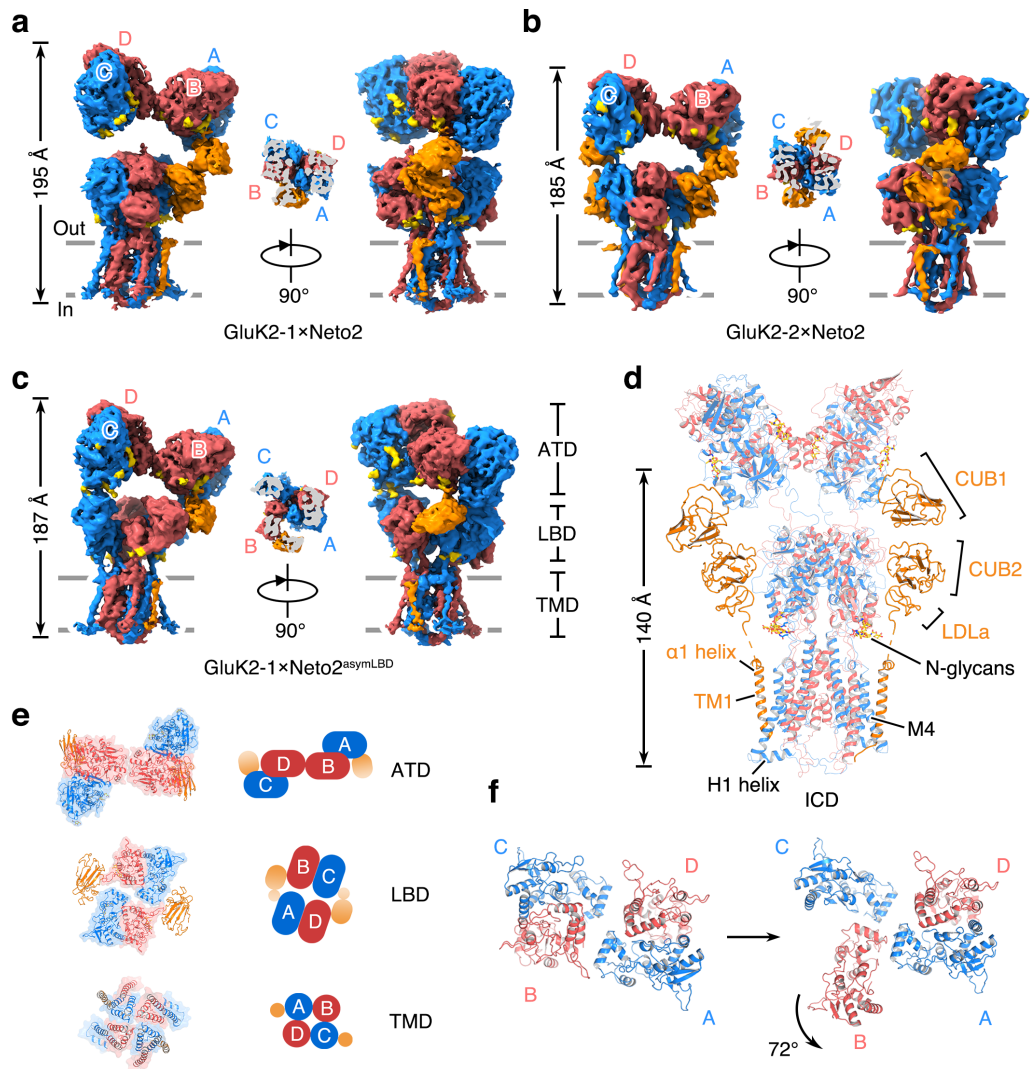

Figure 2

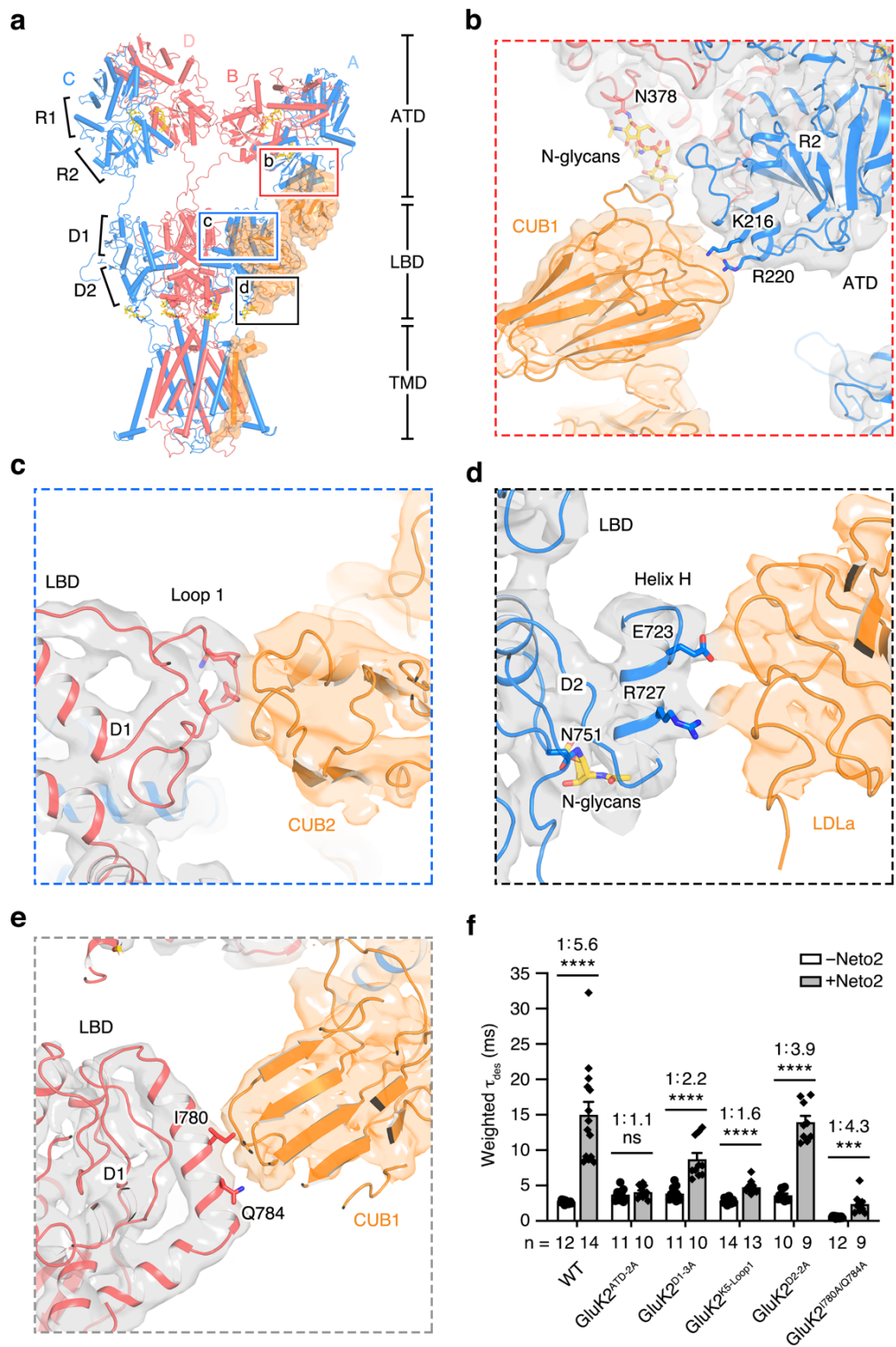

Figure 3

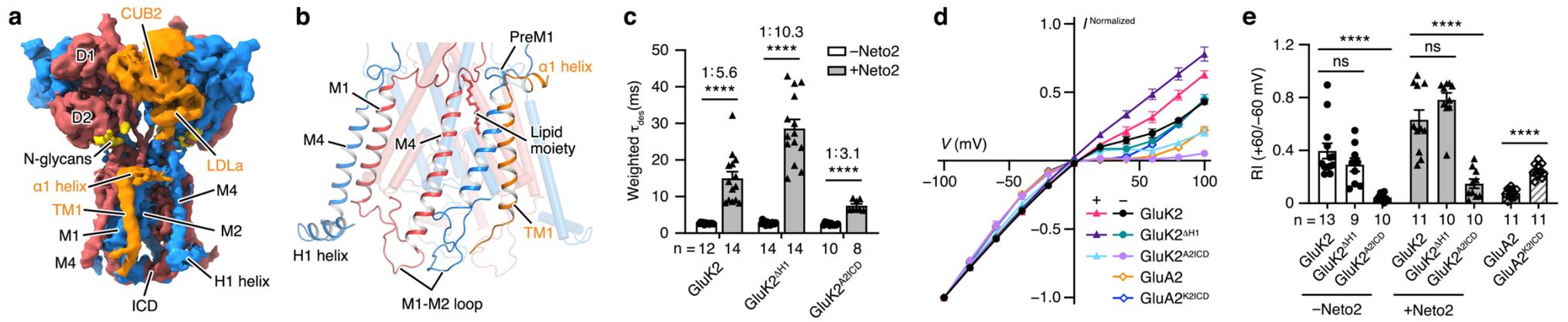

Figure 4

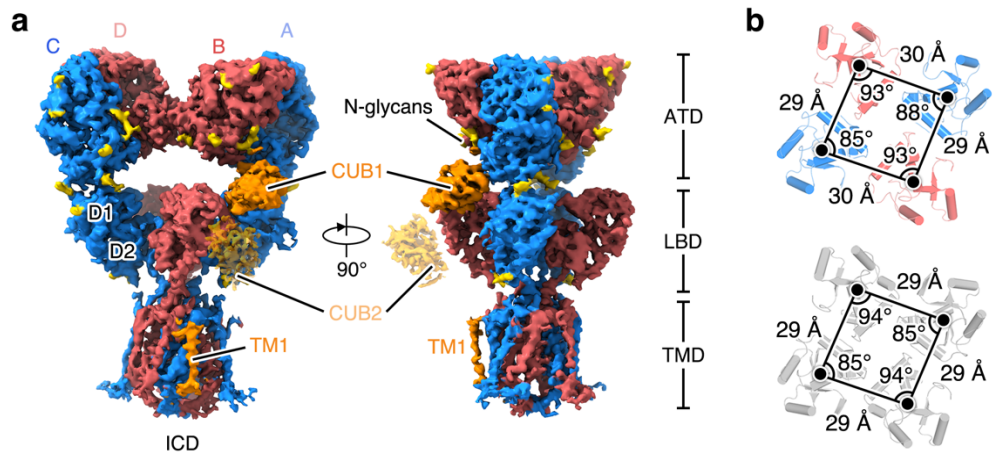

ED Fig. 1

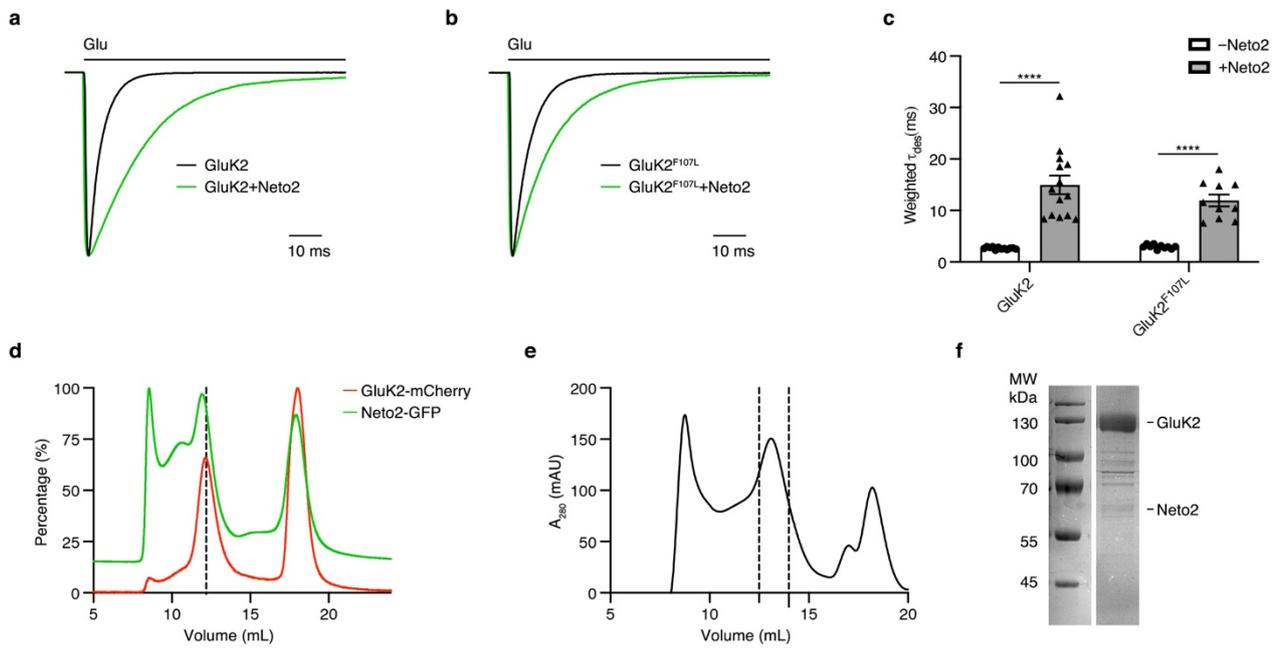

ED Fig. 2

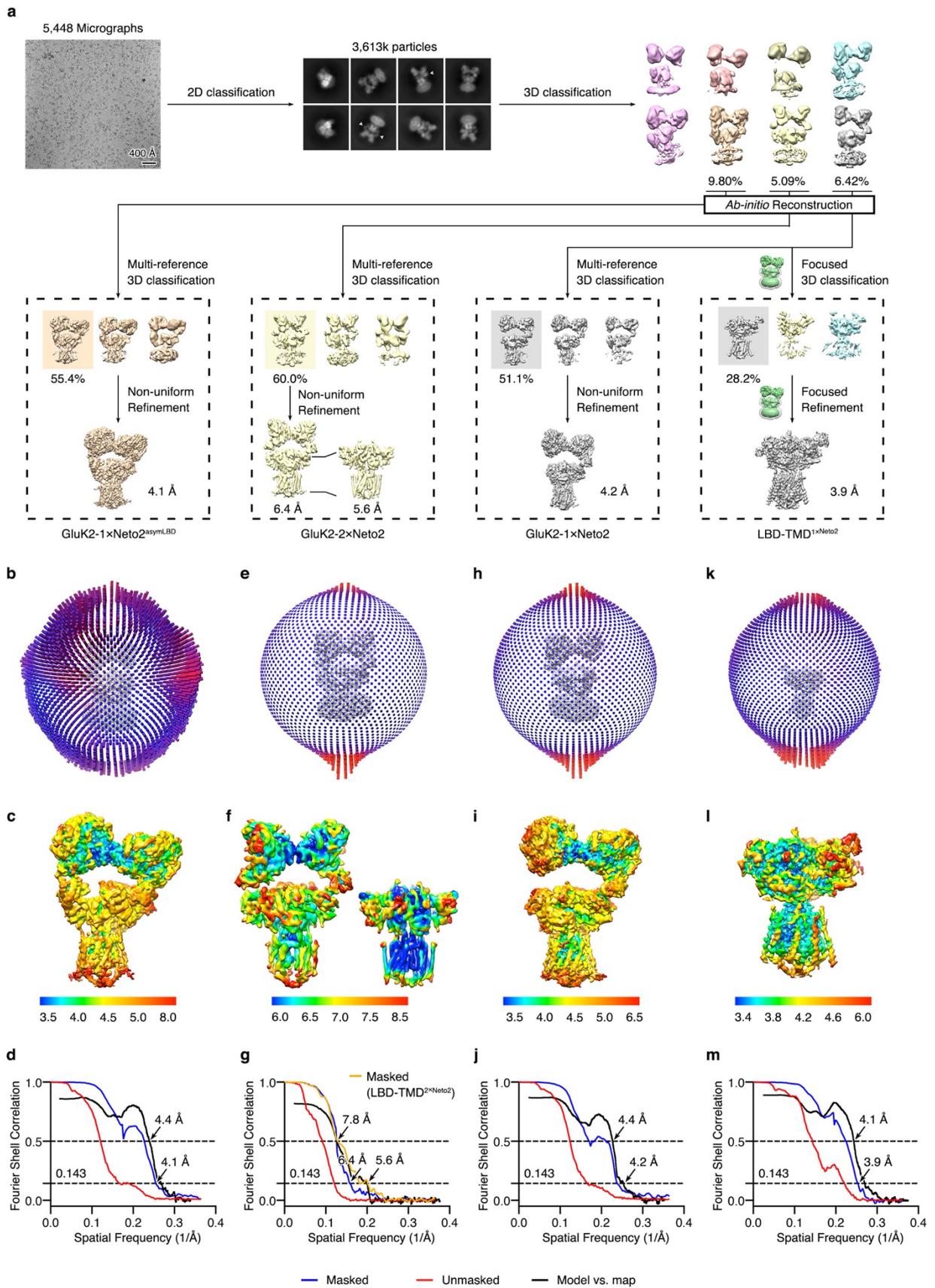

ED Fig. 3

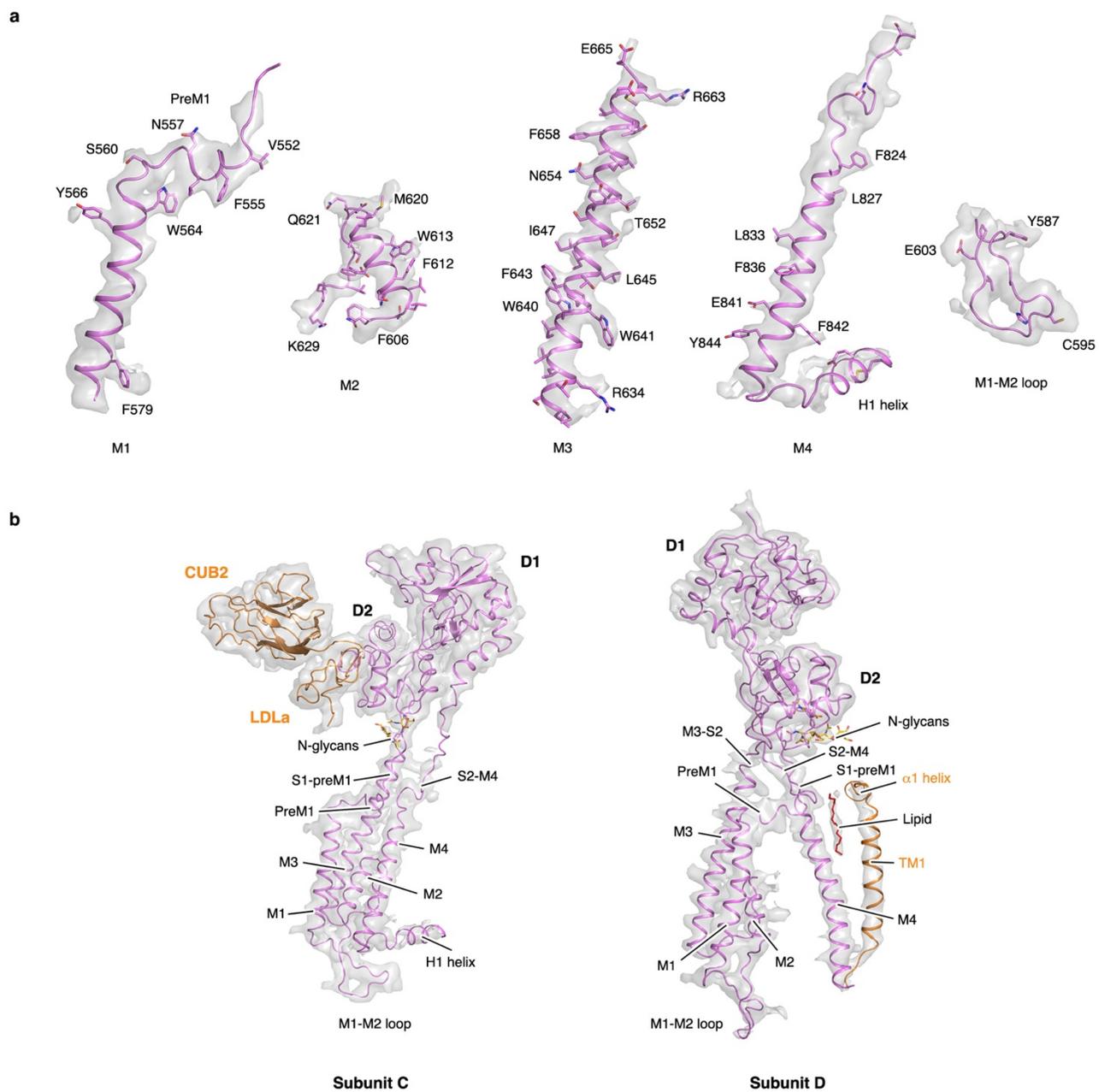

ED Fig. 4

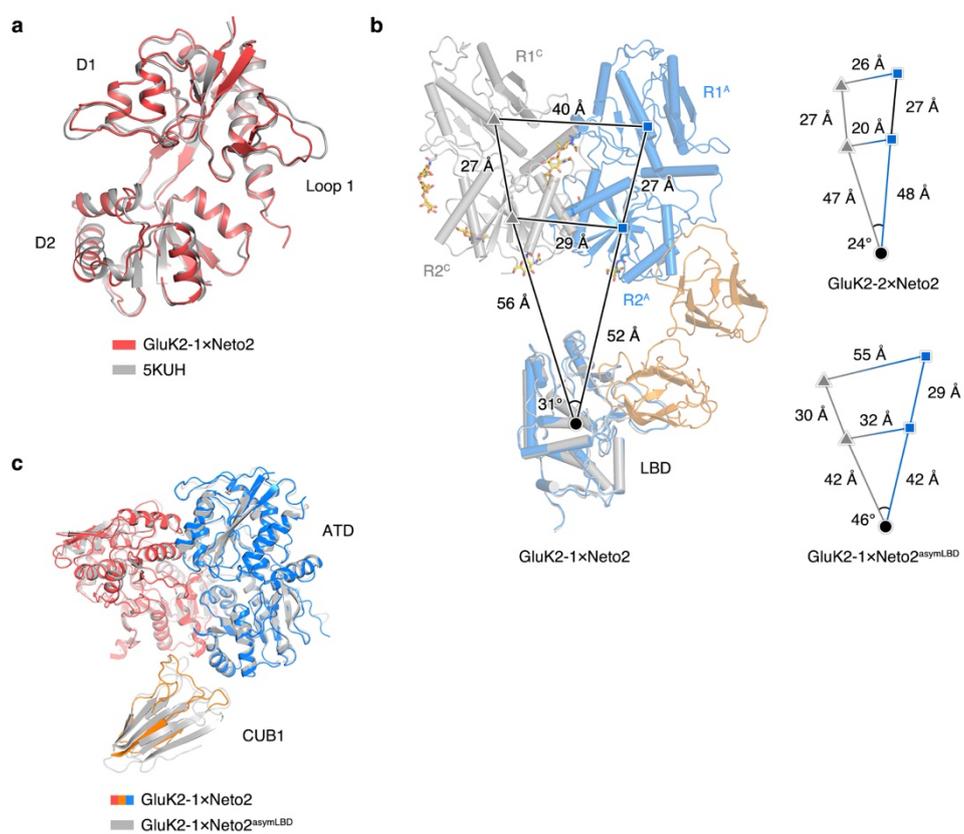

ED Fig. 5

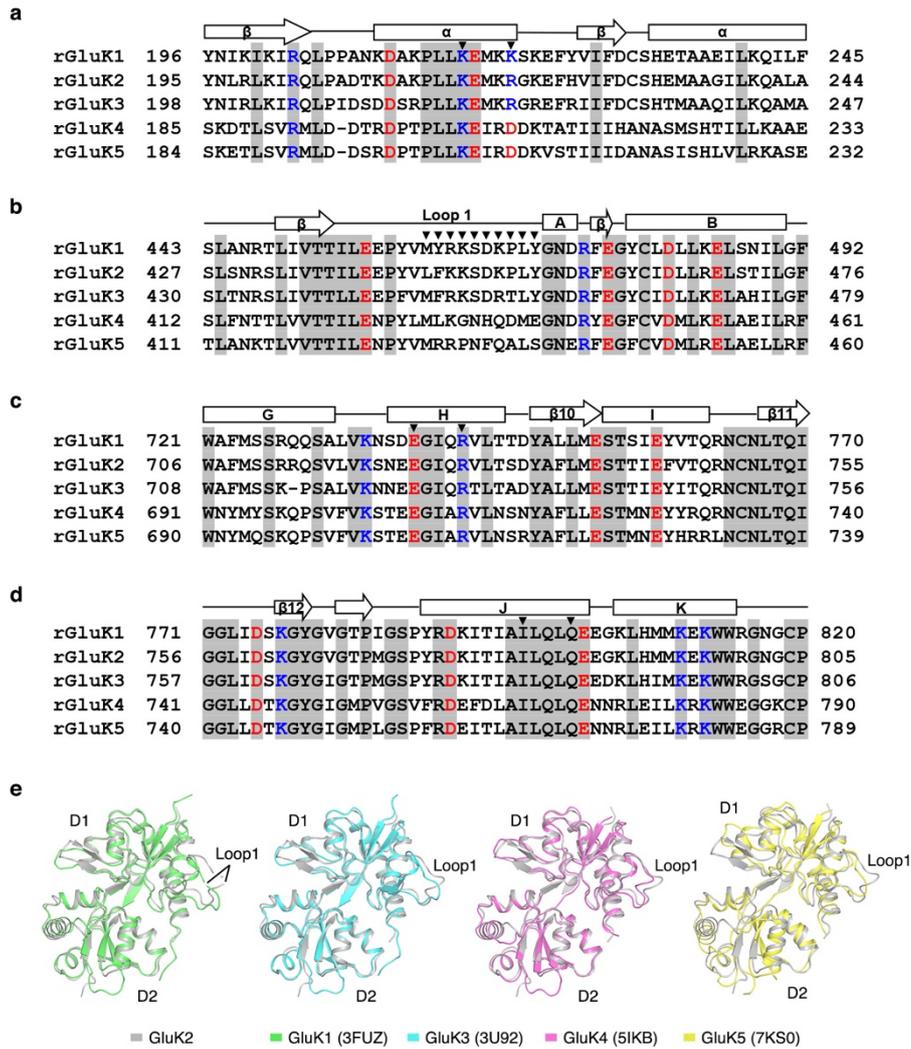

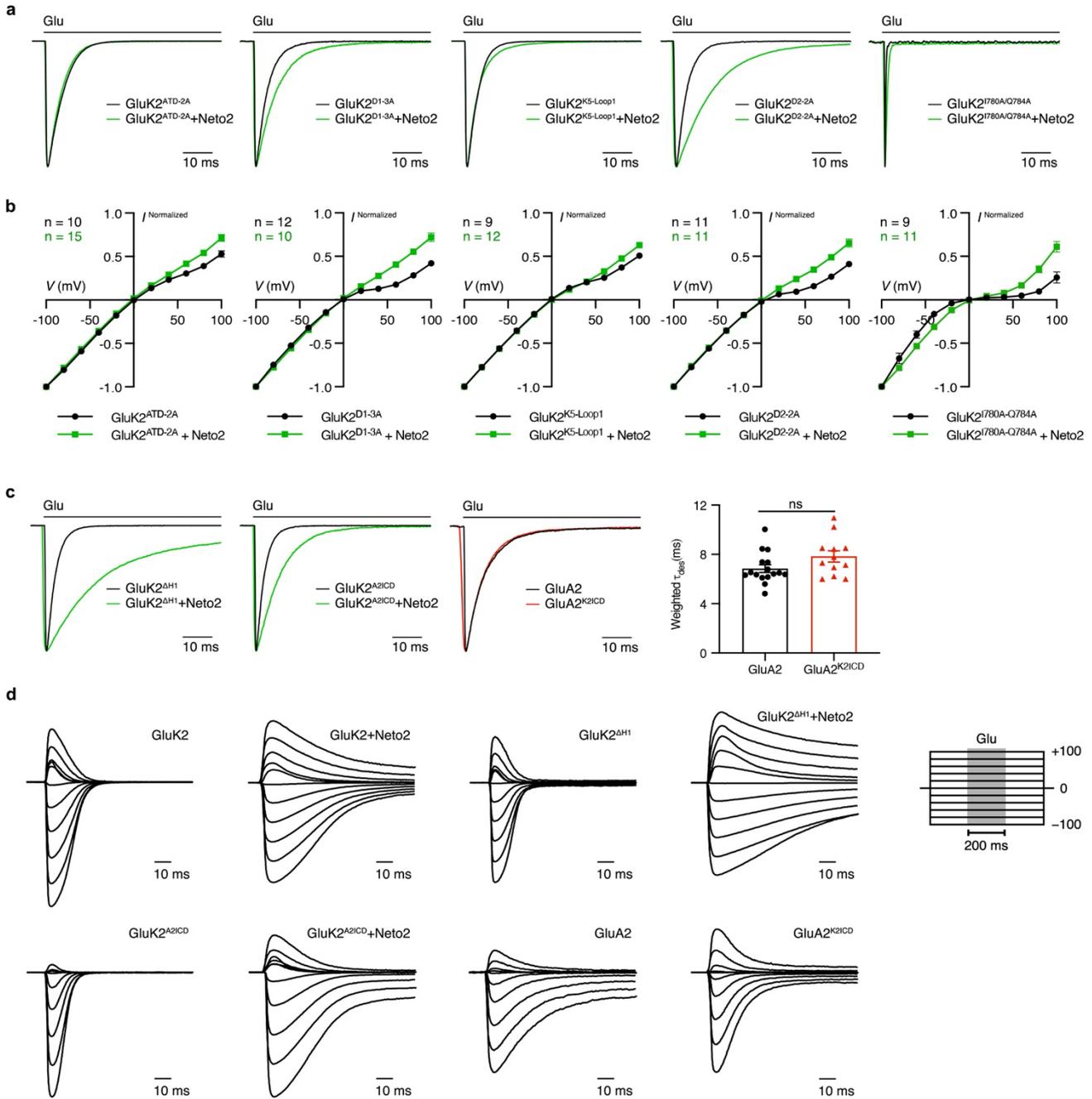

ED Fig. 7

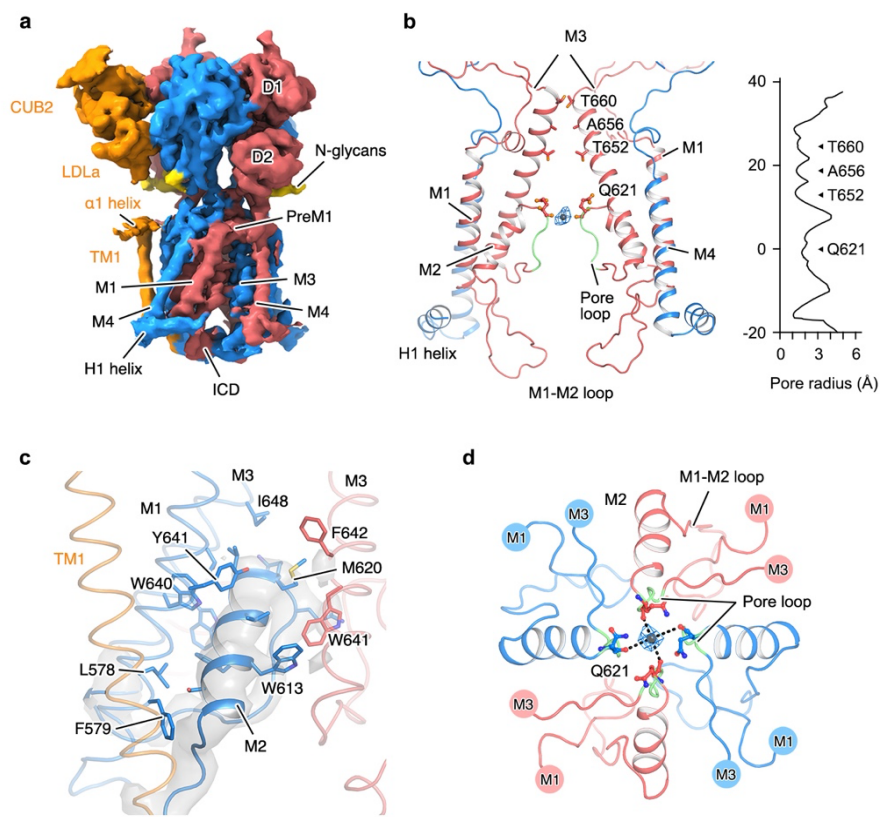

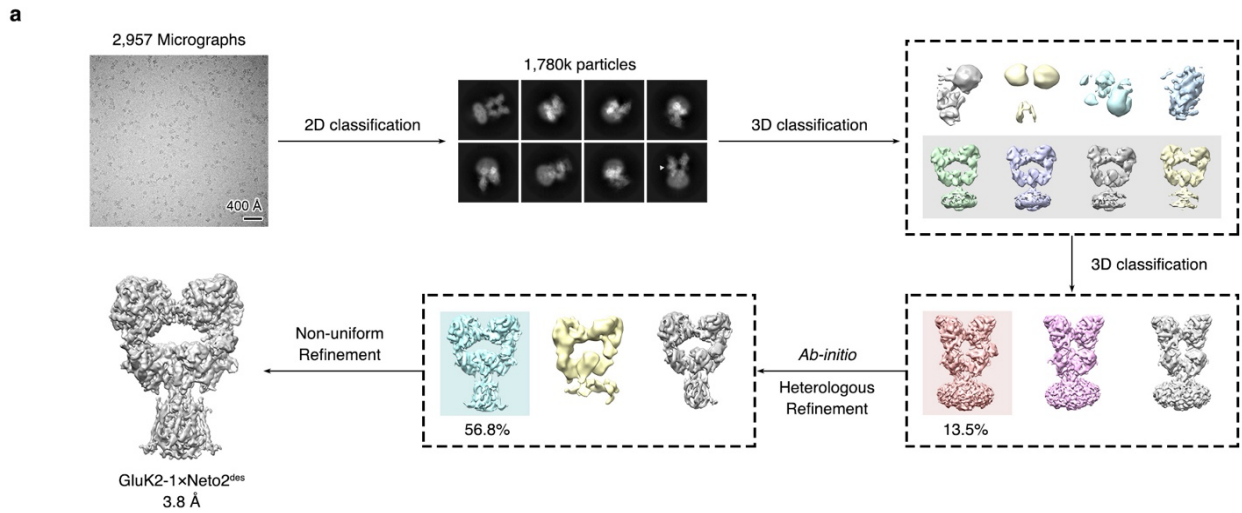
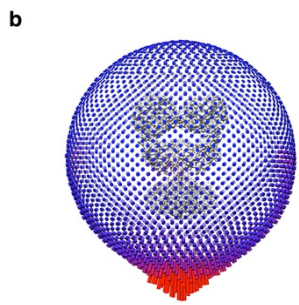
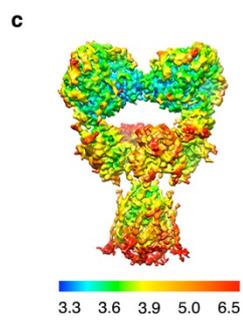
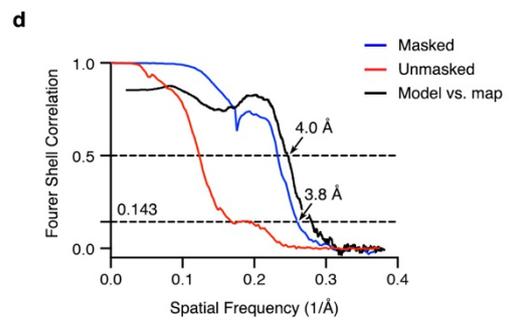

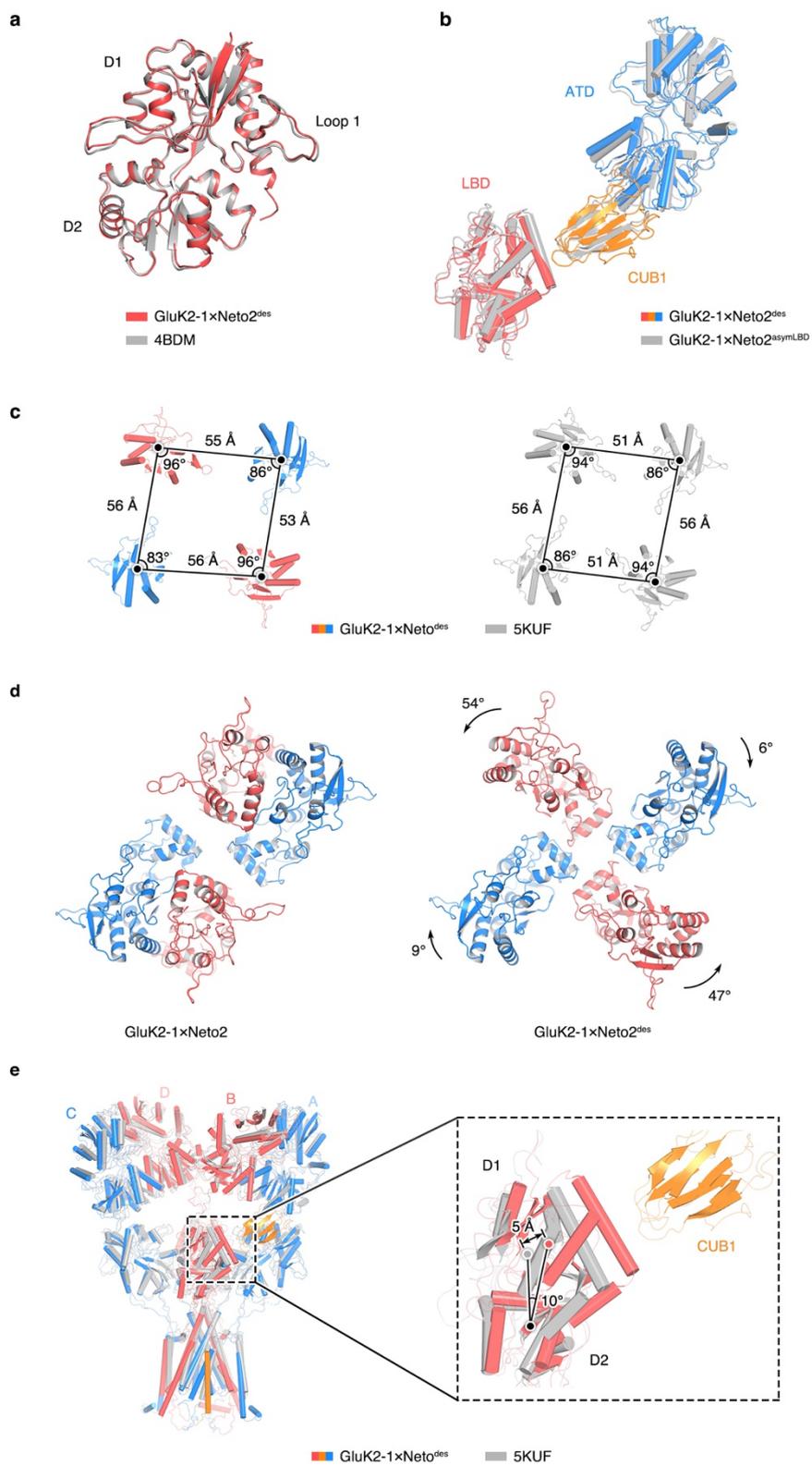

Supplementary Figure 1

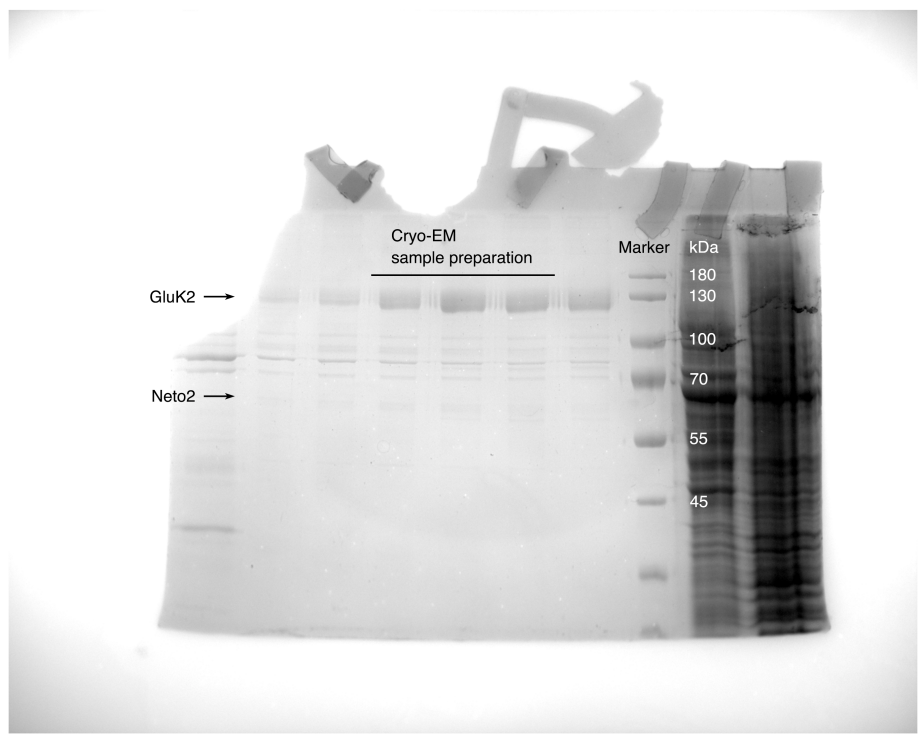